\documentstyle[12pt]{article}
\let\ssection=\section
\renewcommand{\section}{\setcounter{equation}{0}\ssection}

\newcommand\mathC{\mkern1mu\raise2.2pt\hbox{$\scriptscriptstyle|$}
        {\mkern-7mu\rm C}}              % The complex  numbers
\newcommand{\mathR}{{\rm I\! R}}         % The real numbers

\newtheorem{definition}{Definition}[section]

\newcommand\mapdown[1]{\Big\downarrow
                        \rlap{$\vcenter{\hbox{$\scriptstyle#1$}}$}}
\newcommand\mapright[1]{\smash{
        \mathop{\mbox{\large{$\longrightarrow$}}}\limits^{#1}}}
\newcommand\bundle[3]{\begin{array}[t]{c}
        {#1}\\ \mapdown{#2}\\ {#3}\end{array}}
\newcommand\bundlemap[2]{\begin{array}[t]{c}
\mapright{#1}\\
\phantom{\mapdown{}}\\\mapright{#2}\\\end{array}}

\begin{document}
\begin{titlepage}
\hspace{10truecm}Imperial/TP/98--99/76

\begin{center}
{\large\bf Some Possible Roles for Topos Theory in
Quantum Theory and Quantum
Gravity}\footnote{Submitted to {\em Foundations of
Physics} for an issue in honour of Marisa Dalla
Chiara.}\footnote{Based on a lecture given by CJI
at the conference {\em Towards a New Understanding
of Space, Time and Matter\/}, University of
British Columbia, Canada (1999).}
\end{center}

\vspace{0.8 truecm}
\begin{center}
        C.J.~Isham\footnote{email: c.isham@ic.ac.uk}
        \\[10pt]
        The Blackett Laboratory\\
        Imperial College of Science, Technology \& Medicine\\
        South Kensington\\
        London SW7 2BZ\\
\end{center}

\begin{center}
and
\end{center}

\begin{center}
        J.~Butterfield\footnote{email:
            jb56@cus.cam.ac.uk}\\[10pt]
            All Souls College\\
            Oxford OX1 4AL
\end{center}

\begin{center}
   28 September 1999
\end{center}

\begin{abstract}
We discuss some ways in which topos theory (a
branch of category theory) can be applied to
interpretative problems in quantum theory and
quantum gravity. In Section 1, we introduce these
problems. In Section 2, we introduce topos theory,
especially the idea of a topos of presheaves. In
Section 3, we discuss several possible
applications of topos theory to the problems in
Section 1. In Section 4, we draw some conclusions.
\end{abstract}
\end{titlepage}

\section{Introduction}
\label{Sec:Int} In this paper, we wish to suggest
some possible ways in which the notion of a
`topos' can be applied to physics; specifically,
to interpretative problems and foundational issues
in quantum theory and quantum gravity. The first
of these fields is one to which Marisa Dalla
Chiara has contributed so much, especially in its
logical aspects; so it is a pleasure to dedicate
to her a paper focused on logical issues. But the
second field, quantum gravity, also needs to take
cognizance of interpretative problems about
quantum theory; for as we shall describe, research
in quantum gravity soon confronts these problems.
A central theme in this respect is the fundamental
dichotomy in quantum theory between the
traditional instrumentalist interpretation of the
theory, and the essentially realist view of space
and time promulgated by general relativity.
Furthermore, we think there are some significant
ways in which topos theory might be applied in
quantum gravity proper, not all of which are
related directly to the interpretative problems of
quantum theory.

In this Section, we shall introduce these two
fields. In the next Section, we introduce topos
theory, especially the idea of a topos of
presheaves. In Section 3, we briefly discuss
several possible applications of topos theory to
the problems in Section 1; we have developed in
detail elsewhere \cite{IB98,IB99,HIB99} one of
these applications---namely, to the issue of
assigning values to quantum-theoretic quantities.
Finally, in Section 4, we will draw some
conclusions.

\subsection{The Problem of Realism in Quantum Theory}\label{Subsec:IntQuaThe}
Quantum theory has several interpretative
problems, about such topics as measurement and
non-locality; each of which can be formulated in
several ways. But workers in the field would
probably agree that all the problems centre around
the relation between---on the one hand---the
values of physical quantities, and---on the
other---the results of measurement. For our
purposes, it will be helpful to put this in terms
of statements: so the issue is the relation
between ``The quantity $A$ has a value, and that
value is $r$'', (where $r$ is a real number) and
``If a measurement of $A$ is made, the result will
be $r$''.

In classical physics, this relation is seen as
unproblematic. One assumes that, at each moment of
time:
\begin{itemize}
\item[(i)] every physical quantity has a real number as a value
(relative to an appropriate choice of units); and

\item[(ii)] one can measure any
quantity $A$ `ideally', {\em i.e.\/} in such a way
that the result obtained is the value that $A$
possessed before the measurement was made; thus
``epistemology models ontology''.
\end{itemize}

Assumption (i) is implemented mathematically by
the representation of quantities as real-valued
functions on a state space $\Gamma$; so that, in
particular, the statement ``the value of $A$ is
$r$'' ($r\in\mathR$) corresponds to
$\bar{A}^{-1}\{r\}$, the subset of $\Gamma$ that
is the inverse image of the singleton set
$\{r\}\subset\mathR$ under the function
$\bar{A}:\Gamma\rightarrow\mathR$ that represents
the physical quantity $A$. Thus, in particular, to
any state $s \in \Gamma$ there is associated a
`valuation' (an assignment of values) on all
quantities, defined by:
\begin{equation}
V^s(A) := \bar{A}(s).
\end{equation}

 More generally, the proposition ``the value of $A$ is in
 $\Delta$'' (where $\Delta\subset\mathR$) corresponds to the subset
$\bar{A}^{-1}(\Delta)$ of $\Gamma$; these subsets
form a Boolean lattice, which thus provides a
natural representation of the `logic' of
propositions about the system. In particular,
corresponding to the real-numbered valuation $V^s$
on quantities, defined by a state $s \in \Gamma$,
we have a $\{0,1\}$-valued valuation (a
truth-value assignment) to propositions:
\begin{equation}
V^s(A \in \Delta) := 1\mbox{ if }\bar{A}(s) \in
\Delta\mbox{; otherwise }V^s(A \in \Delta) = 0.
\end{equation}
Thus, in particular, in classical physics each
proposition about the system at some fixed time is
regarded as being either true or false.

Note that assumption (ii) is incorporated
implicitly in the formalism---namely, in the
absence of any explicit representation of
measurement---by the fact that the function
$\bar{A}:\Gamma\rightarrow\mathR$ suffices to
represent the quantity $A$, since its values (in
the sense of `values of a function') are the
possessed values (in the sense of `values of a
physical quantity'), and these would be revealed
by an (ideal) measurement.

In quantum theory, on the other hand, the relation
between values and results, and in particular
assumptions (i) and (ii), are notoriously
problematic. The state-space is a Hilbert space
$\cal H$; a quantity $A$ is represented by a
self-adjoint operator $\hat{A}$ (which, with no
significant loss of generality, we can assume
throughout to be bounded), and a statement about
values ``$A \in \Delta$'' corresponds naturally to
a linear subspace of $\cal H$ (or, equivalently,
to a spectral projector, $\hat{E}[A\in\Delta]$, of
$\hat{A}$).

Assumption (i) above (the existence of possessed
values for all quantities) now fails by virtue of
the famous Kochen-Specker theorem \cite{KS67};
which says, roughly speaking, that provided
$\dim({\cal H})>2$, one cannot assign real numbers
as values to all quantum-theory operators in such
a way that for any operator $\hat{A}$ and any
function of it $f(\hat{A})$ ($f$ a function from
$\mathR$ to $\mathR$), the value of $f(\hat{A})$
is the corresponding function of the value of
$\hat{A}$. (On the other hand, in classical
physics, this constraint, called {\em FUNC}, is
trivially satisfied by the valuations $V^s$.) In
particular, it is no longer possible to assign an
unequivocal true-false value to each proposition
of the form ``$A\in\Delta$''.

In a strict instrumentalist approach to quantum
theory, the non-existence of such valuations is of
no great import, since this interpretation of the
theory deals only with the counterfactual
assertion of the probabilities of what values
would be obtained {\em if\/} suitable measurements
are made.

However, strict instrumentalism faces severe
problems (not least in quantum gravity); and the
question arises therefore of whether it may not
after all be possible to retain some `realist
flavour' in the theory by, for example, changing
the logical structure with which propositions
about the values of physical quantities are
handled. One of our claims is that this can indeed
be done by introducing a certain topos perspective
on the Kochen-Specker theorem.

We will argue for this claim in Section 3.7. For
the moment, we just remark that no-go theorems
like that of Kochen and Specker depend upon the
fact that the set of all spectral projectors of
$\cal H$ form a non-Boolean, indeed
non-distributive, lattice; suggesting a
non-Boolean, indeed non-distributive, `quantum
logic'. This alluring idea, originated by Birkhoff
and von Neumann \cite{BVN36}, has been greatly
developed in various directions.\footnote{Cf.
Dalla Chiara and Giuntini's masterly recent survey
\cite{DCG99}. This survey includes recent
developments that generalize the basic
correspondence between subspaces and propositions
about values, so as to treat so-called `unsharp'
(`operational') quantum physics; on this see also
\cite{DCG94} and other papers in this issue.} But
in this connection, the important point to stress
for the purposes of this paper is that the logic
associated with our topos-theoretic proposals is
not non-distributive. On the contrary, {\em any\/}
topos has an associated internal logical structure
that {\em is\/} distributive. This retention of
the distributive law marks a major departure from
the dominant tradition of quantum logic stemming
from Birkhoff and von Neumann.

On the other hand, our proposals do involve
non-Boolean structure since the internal logic of
a topos is `intuitionistic', in the sense that the
law of excluded middle may not hold (although for
some toposes, such as the category of sets, it
does apply).\footnote{Some intuitionistic
structures also arise in the dominant
`non-distributive' tradition in quantum logic; for
example, in the Brouwer-Zadeh approach to unsharp
quantum theory; cf. \cite{CL94}.}

\subsection{Challenges of Quantum Gravity}\label{Subsec:ChaQuaGra}
The problem of realism becomes particularly acute
in the case of quantum gravity. This field is
notoriously problematic in comparison with other
branches of theoretical physics, not just
technically but also conceptually. In the first
place, there is no clear agreement about what the
aim of a quantum theory of gravity should be,
apart from the broad goal of in some way unifying,
or reconciling, quantum theory and general
relativity. That these theories do indeed conflict
is clear enough: general relativity is a highly
successful theory of gravity and spacetime, which
treats matter classically (both as a source of the
gravitational field, and as influenced by it) and
treats the structure of spacetime as dynamical;
while quantum theory provides our successful
theories of matter, and treats spacetime as a
fixed, background structure.

Much has been written about the conceptual
problems that arise in quantum gravity; (for one
recent survey, cf. \cite{BI99}). But in the
present context it suffices to say that these are
sufficiently severe to cause a number of workers
in the field to question many of the basic ideas
that are implicit in most, if not all, of the
existing programmes. For example, there have been
a number of suggestions that spatio-temporal ideas
of classical general relativity such as
topological spaces, continuum manifolds,
space-time geometry, micro-causality, etc.\ are
inapplicable in quantum gravity.

More iconoclastically, one may doubt the
applicability of quantum theory itself,
notwithstanding the fact that all current research
programmes in quantum gravity do adopt a
more-or-less standard approach to quantum theory.
In particular, as we shall discuss shortly, there
is a danger of certain {\em a priori\/}, classical
ideas about space and time being used unthinkingly
in the very formulation of quantum theory; thus
leading to a type of category error when attempts
are made to apply this theory to domains in
quantum gravity where such concepts may be
inappropriate.

\subsection{Whence the Continuum?}
\label{Subsec:WheConMod} As an example of the
adoption by quantum theory of certain problematic
concepts, we will now consider the use of the
continuum---{\em i.e.}, of real and complex
numbers---in the formulation of our physical
theories in general. And having raised this topic,
we shall describe in the next Subsection two
natural alternative conceptions of space and time,
which will involve the use of topos theory. (We
give this discussion before introducing toposes in
Section 2, since: (i) it is independent of the
logical issues that will be emphasised in the rest
of this paper; and accordingly, (ii) it can be
understood without using details of the notion of
a topos.)

So let us ask: why do we use the continuum, {\em
i.e.}, the real numbers, in our physical theories?
The three obvious answers are: (i) to be the
values of physical quantities; (ii) to model space
and time; and (iii) to be the values of
probabilities. But let us pursue a little the
question of what justifies these answers: we will
discuss them in turn.
\begin{itemize}
\item {As to (i), the first point to recognize is of course that the whole
edifice of physics, both classical and quantum,
depends upon applying calculus and its higher
developments (for example, functional analysis and
differential geometry) to the values of physical
quantities. But in the face of this, one could
still take the view that the success of these
physical theories only shows the `instrumental
utility' of the continuum---and not that physical
quantities really have real-number values. This is
not the place to enter the general philosophical
debate between instrumentalist and realist views
of scientific theories; or even the more specific
question of whether an instrumentalist view about
the continuum is committed to somehow rewriting
all our physical theories without use of $\mathR$:
for example, in terms of rational numbers (and if
so, how he should do it!). Suffice it to say here
that the issue whether physical quantities have
real-number values leads into the issue whether
space itself is modelled using $\mathR$. For not
only is length one (obviously very important!)
quantity in physics; also, one main, if not
compelling, reason for taking other quantities to
have real-number values is that results of
measuring them can apparently always be reduced to
the position of some sort of pointer in
space---and space is modelled using $\mathR$.

We note that the formalism of elementary wave
mechanics affords a good example of an {\em a
priori\/} adoption of the idea of a continuum
model of space: indeed, the $x$ in $\psi(x)$
represents space, and in the theory this
observable is modelled as having a continuous
spectrum; in turn, this requires the underlying
Hilbert space to be defined over the real or
complex field. }

\item So we turn to (ii): why should space be modelled using $\mathR$?
More specifically, we ask, in the light of our
remarks about (i): Can any reason be given apart
from the (admittedly, immense) `instrumental
utility' of doing so, in the physical theories we
have so far developed? In short, our answer is No.
In particular, we believe there is no good {\em a
priori\/} reason why space should be a continuum;
similarly, {\em mutatis mutandis\/} for time. But
then the crucial question arises of how this
possibility of a non-continuum space should be
reflected in our basic theories, in particular in
quantum theory itself, which is one of the central
ingredients of quantum gravity.

\item As to (iii), why should probabilities be real numbers?
Admittedly, if probability is construed in terms
of the relative frequency of a result in a
sequence of measurements, then real numbers do
arise as the limits of infinite sequences of
finite relative frequencies (which are all
rational numbers). But this limiting relative
frequency interpretation of probability is
disputable. In particular, it seems problematic in
the quantum gravity regime where standard ideas of
space and time might break down in such a way that
the idea of spatial or temporal `ensembles' is
inappropriate.

On the other hand, for the other main
interpretations of probability---subjective,
logical, or propensity---there seems to us to be
no compelling {\em a priori\/} reason why
probabilities should be real numbers. For
subjective probability (roughly: what a rational
agent's minimum acceptable odds, for betting on a
proposition, are or should be): many authors point
out that the use of $\mathR$ as the values of
probabilities is questionable, whether as an
idealization of the psychological facts, or as a
norm of rationality. For the logical and
propensity interpretations---which are arguably
more likely to be appropriate for the quantum
gravity regime---the use of $\mathR$ as the values
of probabilities is less discussed. But again, we
see no {\em a priori\/} reason for
$\mathR$.\footnote{It seems to us that in the
literature, the principal `justification' given
for $\mathR$ is the mathematical desideratum of
securing a uniqueness claim in a representation
theorem about axiom systems for qualitative
probability; the claim is secured by imposing a
continuity axiom that excludes number-fields other
than $\mathR$ as the codomain of the representing
probability-function.} Indeed, we would claim that
while no doubt in some cases, one `degree of
entailment' or `propensity' is `larger' than
another, it also seems possible that in other
cases two degrees of entailment, or two
propensities, might be incomparable--so that the
codomain of the probability-function should be,
not a linear order, but some sort of partially
ordered set (equipped with a sum-operation, so as
to make sense of the additivity axiom for
probabilities). Once again this suggests that a
fairly radical revision of quantum theory itself
might be in order.
\end{itemize}

\subsection{Alternative Conceptions of Spacetime}
Scepticism about the use of the continuum in
present-day physical theories prompts one to
consider alternative conceptions of space and
time. We turn to briefly sketch two such
conceptions. Both involve topos theory, and indeed
raise the idea---even more iconoclastic than
scepticism about the continuum---
 that the use of set theory itself may be inappropriate for
modelling space and time.

\paragraph{1.4.1 From points to regions} \label{Para:Loc}
In standard general relativity---and, indeed, in
all classical physics---space (and similarly time)
is modelled by a set, and the elements of that set
are viewed as corresponding to points in space.
However, if one is `suspicious of
points'---whether of spacetime, of space or of
time ({\em i.e.\/} instants)---it is natural to
try and construct a theory based on `regions' as
the primary concept; with `points'---if they exist
at all---being relegated to a secondary role in
which they are determined by the `regions' in some
way (rather than regions being sets of points, as
in the standard theories).\footnote{For time, the
natural word is `intervals', not `regions'; but we
shall use only `regions', though the discussion to
follow applies equally to the one-dimensional
case---and so to time---as it does to
higher-dimensional cases, and so to space and
spacetime.}

So far as we know, the first rigorous development
of this idea was made in the context of
foundational studies in the 1920s and 1930s, by
authors such as Tarski. The idea was to write down
axioms for regions from which one could construct
points, with the properties they enjoyed in some
familiar theory such as three-dimensional
Euclidean geometry. For example, the points were
constructed in terms of sequences of regions, each
contained in its predecessor, and whose `widths'
tended to zero; (more precisely, the point might
be identified with an equivalence class of such
sequences). The success of such a construction was
embodied in a representation theorem, that any
model of the given axiom system for regions was
isomorphic to, for example, $\mathR^3$ equipped
with a structured family of subsets, which
corresponded to the axiom system's regions. In
this sense, this line of work was `conservative':
one recovered the familiar theory with its points,
from a new axiom system with regions as
primitives.\footnote{From the pure mathematical
point of view, Stone's representation theorem for
Boolean algebras of 1936 was a landmark for this
sort of work.}

But use of regions in place of points need not be
`conservative': one can imagine axiom systems for
regions, whose models (or some of whose models) do
not contain anything corresponding to points of
which the regions are composed. Indeed, for any
topological space $Z$, the family of all open sets
can have algebraic operations of `conjunction',
`disjunction' and `negation' defined on them by:
$O_1\land O_2:=O_1\cap O_2$; $O_1\lor O_2:=O_1\cup
O_2$; and $\neg O:={\rm int}(Z-O)$; and with these
operations, the open sets form a complete Heyting
algebra, also known as a {\em locale\/}. Here, a
Heyting algebra is defined to be a distributive
lattice $H$, with null and unit elements, that is
{\em relatively complemented\/}, which means that
to any pair $S_1,S_2$ in $H$, there exists an
element $S_1\Rightarrow S_2$ of $H$ with the
property that, for all $S\in H$,
\begin{equation}
        S\leq (S_1\Rightarrow S_2)\mbox{  if and only if
$S\land S_1\leq S_2$}.
\end{equation}
Heyting algebras are thus a generalization of
Boolean algebras; they need not obey the law of
excluded middle, and so provide natural algebraic
structures for intuitionistic logic. A Heyting
algebra is said to be {\em complete\/} if every
family of elements has a least upper bound.
Summing up: the open sets of any topological space
form a Heyting algebra, when partially ordered by
set-inclusion; indeed a complete Heyting algebra
(a locale), since arbitrary unions of open sets
are open.

 However, it turns out that not every locale is isomorphic to the
Heyting algebra of open sets of some topological
space; and in this sense, the theory of regions
given by the definition of a locale is not
`conservative'---it genuinely generalizes the idea
of a topological space, allowing families of
regions that are not composed of underlying
points.

A far-reaching generalisation of this idea is
given by topos theory. As we shall see in Section
2.2: (i) in any topos, there is an analogue of the
set-theoretic idea of the family of subsets of a
given set---called the family of subobjects of a
given object $X$; (ii) for any object $X$ in any
topos, the family of subobjects of $X$ is a
locale.

\paragraph{1.4.2 Synthetic Differential Geometry}
\label{Para:SynDifGeo} Recent decades have seen a
revival of the idea of infinitesimals. Though the
idea was heuristically valuable in the discovery
and development of the calculus, it was expunged
in the nineteenth-century rigorization of analysis
by authors such as Cauchy and Weierstrass---for
surely no sense could be made of the idea of
nilpotent real numbers, {\em i.e.}, $d$ such that
$d^2 = 0$, apart from the trivial case $d = 0$?
But it turns out that sense {\em can\/} be made of
this: indeed in two somewhat different ways.

In the first approach, called `non-standard
analysis', every infinitesimal ({\em i.e.}, every
nilpotent $d \neq 0$) has a reciprocal, so that
there are different infinite numbers corresponding
to the different infinitesimals. There were
attempts in the 1970s to apply this idea to
quantum field theory: in particular, it was shown
how the different orders of ultra-violet
divergences that arise correspond to different
types of infinite number in the sense of
non-standard analysis \cite{Far75}.

However, we wish here to focus on the alternative
approach in which we have infinitesimals, but
without the corresponding infinite numbers. It
transpires that this is possible provided we work
within the context of a topos; for example, a
careful study of the proof that the only real
number $d$ such that $d^2=0$ is $0$, shows that it
involves the principle of excluded middle, which
in general does not hold in the characteristic
intuitionistic logic of a topos \cite{Lav96}.

So in this second approach, called `synthetic
differential geometry', infinitesimals do not have
reciprocals. Applying this approach to elementary
real analysis, `all goes smoothly'! For example,
all functions are differentiable, with the linear
approximation familiar from Taylor's theorem, $f(x
+ d) = f(x) + d\,f'(x)$, being exact. And in the
context of synthetic differential geometry, a
tangent vector on a manifold $\cal M$ is a map
(more precisely, a `morphism') from the object
$D:= \{d \mid d^2 = 0 \}$ to $\cal M$.

Furthermore, one can go on to apply this approach
to the higher developments of calculus. Indeed,
this has already been done by mathematicians; but
we shall not try to report, let alone sketch, any
such applications.

One crucial question is whether or not there are
any {\em physically\/} natural applications of
synthetic differential geometry to physics; (as
against `merely rewriting' standard theories in
synthetic terms). We will claim in Section 3 that
precisely such an application arises in the
consistent-histories formulation of quantum theory
in the context of a continuous time variable.

\section{Presheaves and Related Notions from Topos Theory}
\label{Sec:PreRelNotfroTop} There are various
approaches to the notion of a topos but we will
focus here on one that emphasises the underlying
logical structure (as befits a Festschrift for
Marisa Dalla Chiara!) Also, to keep the discussion
simple, we will not develop the full definition of
a topos---which our discussions of applications in
Section 3 will in fact not need. Indeed, in this
Section we will only discuss one, albeit crucial,
clause of the definition of a topos: the
requirement that a topos contain a `subobject
classifier'. This is a generalization of the idea,
familiar in set-theory, of characteristic
functions. The generalization will turn out to
have a particularly interesting logical structure
in the case of the kind of topos to which our
discussion in Section 3 is confined: a topos of
presheaves.

A topos is a particular type of category. Very
roughly, it is a category that behaves much like
the category of sets; indeed, this category, which
we will call ${\rm Set}$, is itself a topos. So we
will begin by recalling a few fundamental concepts
that apply to any category (Section 2.1); then we
will discuss the idea of a subobject classifier
(Section 2.2); and finally, the ideas of a
presheaf, and a topos of presheaves (Section 2.3).

\subsection{Categories}\label{Subsec:Cat}
We recall that a category consists of a collection
of {\em objects\/}, and a collection of {\em
arrows\/} (or {\em morphisms\/}), with the
following three properties. (1) Each arrow $f$ is
associated with a pair of objects, known as its
{\em domain\/} (dom $f$) and the {\em codomain\/}
(cod $f$), and is written in the form
$f:B\rightarrow A$ where $B ={\rm dom} f$ and
$A={\rm cod} f$. (2) Given two arrows
$f:B\rightarrow A$ and $g:C\rightarrow B$ (so that
the codomain of $g$ is equal to the domain of
$f$), there is a composite arrow $f\circ
g:C\rightarrow A$; and this composition of arrows
obeys the associative law. (3) Each object $A$ has
an identity arrow, ${\rm id}_A:A\rightarrow A$,
with the properties that for all $f:B\rightarrow
A$ and all $g:A\rightarrow C$, ${\rm id}_A\circ f
= f$ and $g\circ {\rm id}_A = g$.

We have already mentioned the prototype category
(indeed, topos) ${\rm Set}$, in which the objects
are sets and the arrows are ordinary functions
between them (set-maps). In many categories, the
objects are sets equipped with some type of
additional structure, and the arrows are functions
that preserve this structure (hence the word
`morphism'). An obvious algebraic example is the
category of groups, where an object is a group,
and an arrow $f:G_1\rightarrow G_2$ is a group
homomorphism from $G_1$ to $G_2$. (More generally,
one often defines one category in terms of
another; and in such a case, there is often only
one obvious way of defining composition and
identity maps for the new category.) However, a
category need not have `structured sets' as its
objects. An example (which will be prominent in
Section 3) is given by any partially-ordered set
(`poset') $\cal P$. It can be regarded as a
category in which (i) the objects are the elements
of $\cal P$; and (ii) if $p,q\in\cal P$, an arrow
from $p$ to $q$ is defined to exist if, and only
if, $p\leq q$ in the poset structure. Thus, in a
poset regarded as a category, there is at most one
arrow between any pair of objects $p,q\in\cal P$.

In any category, an object $T$ is called {\em a
terminal\/} (resp.\ {\em initial\/}) object if for
every object $A$ there is exactly one arrow
$f:A\rightarrow T$ (resp.\ $f:T\rightarrow A$).
Any two terminal (resp. initial) objects are
isomorphic\footnote{Two objects $A$ and $B$ in a
category are said to be {\em isomorphic\/} if
there exists arrows $f:A\rightarrow B$ and
$g:B\rightarrow A$ such that $f\circ g={\rm id}_B$
and $g\circ f={\rm id}_A$}. So we normally fix on
one such object; and we write `the' terminal
(resp.\ initial) object as ${\bf 1}$ (resp. ${\bf
0}$). An arrow ${\bf 1}\rightarrow A$ is called a
{\em point\/}, or a {\em global element\/}, of
$A$. For example, applying these definitions to
our example ${\rm Set}$ of a category, we find
that (i) each singleton set is a terminal object;
(ii) the empty set $\emptyset$ is initial; and
(iii) the points of $A$ give a `listing' of the
elements of $A$.

\subsection{Toposes and Subobject Classifiers}
\label{Subsec:TopSubCla} We turn now to
introducing a very special kind of category called
a `topos'. As we said at the beginning of this
Section, we will discuss only one clause of the
definition of a topos: the requirement that a
topos contain a generalization of the
set-theoretic concept of a characteristic
function; this generalization is closely related
to what is called a `subobject classifier'.

Recall that characteristic functions classify
whether an element $x$ is in a given subset $A$ of
a set $X$ by mapping $x$ to $1$ if $x \in A$, and
to $0$ if $x \notin A$. More fully: for any set
$X$, and any subset $A \subseteq X$, there is a
characteristic function
$\chi_A:X\rightarrow\{0,1\}$, with $\chi_{A}(x) =
1$ or 0 according as $x \in A$ or $x \notin A$.
One thinks of $\{0,1\}$ as the truth-values; and
$\chi_A$ classifies the various $x$ for the
set-theoretically natural question, ``$x \in
A$?''. Furthermore, the structure of ${\rm
Set}$---the category of sets---secures the
existence of this set of truth-values and the
various functions $\chi_A$: in particular,
$\{0,1\}$ is itself a set, {\em i.e.\/} an object
in the category ${\rm Set}$, and for each $A, X$
with $A \subseteq X$, $\chi_A$ is an arrow from
$X$ to $\{0,1\}$.

It is possible to formulate this `classifying
action' of the various $\chi_A$ in general
category-theoretic terms, so as to give a fruitful
generalization. For the purposes of this paper,
the main ideas are as follows.
\begin{enumerate}
\item In any category, one can define
 a categorial analogue of the set-theoretic idea of
subset: it is called a `subobject'. More
precisely, one generalizes the idea that a subset
$A$ of $X$ has a preferred injective ({\em
i.e.\/}, one-to-one) map $A \rightarrow X$ sending
$x \in A$ to $x \in X$. For category theory
provides a generalization of injective maps,
called `monic arrows' or `monics'; so that in any
category one defines a subobject of any object $X$
to be a monic with codomain $X$.

\item Any topos is required to have an analogue, written $\Omega$,
of the set $\{0,1\}$ of truth-values. That is to
say: just as $\{0,1\}$ is itself a set---{\em
i.e.}, an object in the category ${\rm Set}$ of
sets---so also in any topos, $\Omega$ is an object
in the topos. And just as the set of subsets of a
given set $X$ corresponds to the set of
characteristic functions from subsets of $X$ to
$\{0,1\}$; so also in any topos, there is a
one-to-one correspondence between subobjects of an
object $X$, and arrows from $X$ to $\Omega$.

\item {In a topos, $\Omega$ acts as an object of generalized truth-values,
just as $\{0,1\}$ does in set-theory; (though
$\Omega$ typically has more than two global
elements). Intuitively, the elements of $\Omega$
are the answers to a natural `multiple-choice
question' about the objects in the topos, just as
``$x \in X$?'' is natural for sets. An example:

\begin{itemize}
\item {A set $X$ equipped with a given function $\alpha:X\rightarrow X$ is called
an {\em endomap\/}, written $(X;\alpha)$; and the
family of all endomaps forms a category---indeed,
a topos---when one defines an arrow from
$(X;\alpha)$ to $(Y;\beta)$ to be an ordinary
set-function $f$ between the underlying sets, from
$X$ to $Y$, that preserves the endomap structure,
{\em i.e.}, $f \circ \alpha = \beta \circ f$.

Applying the definition of a subobject, it turns
out that a subobject of $(X;\alpha)$ is a subset
of $X$ that is closed under $\alpha$, equipped
with the restriction of $\alpha$: {\em i.e.}, a
subobject is $(Z,\alpha\mid_Z)$, with $Z \subseteq
X$ and such that $\alpha(Z)\subset Z$. So a
natural question, given $x \in X$ and a subendomap
$(Z,\alpha\mid_Z)$, is: ``How many iterations of
$\alpha$ are needed to send $x$ (or rather its
descendant, $\alpha(x)$ or $\alpha^2(x)$ or
$\alpha^3(x)$ \ldots) into $Z$?'' The possible
answers are `$0$ ({\em i.e.}, $x\in Z$)', `$1$',
`$2$',\ldots, and `infinity ({\em i.e.}, the
descendants never enter $Z$)'; and if the answer
for $x$ is some natural number $N$ (resp.\ $0$,
infinity), then the answer for $\alpha(x)$ is
$N-1$ (resp.\ $0$, infinity). So the possible
answers can be presented as an endomap, with the
elements of the base-set labelled as `0', `1',
`2', ..., and `$\infty$', and with the map
$\alpha$ acting as follows: $\alpha:N \mapsto N-1$
for $N = 1,2,...$, and $\alpha: 0 \mapsto 0$,
$\alpha:\infty \mapsto \infty$.

And it turns out that this endomap is exactly the
object $\Omega$ in the category of endomaps!
Recall that in any topos $\Omega$ is an object in
the topos, so that here $\Omega$ must itself be an
endomap, a set equipped with a function to
itself.}
\end{itemize}
}

\item  This example suggests that $\Omega$ is fixed by the
structure of the topos concerned. And indeed, this
is so in the precise sense that, although the
clause in the definition of a topos that
postulates the existence of $\Omega$ characterizes
$\Omega$ solely in terms of conditions on the
topos' objects and arrows, $\Omega$ is provably
unique (up to isomorphism).

Furthermore, in any topos, $\Omega$ has a natural
logical structure. More exactly, $\Omega$ has the
internal structure of a Heyting algebra object:
the algebraic structure appropriate for
intuitionistic logic, mentioned in Section 1.4.1.
In addition, in any topos, the collection of
subobjects of any given object $X$ is a complete
Heyting algebra (a locale). We shall see this sort
of Heyting algebra structure in more detail in the
next Subsection, for the case that concerns
us---presheaves. For the moment we note only
 the general point, valid for any topos, that because
$\Omega$ is fixed by the structure of the topos
concerned, and has a natural Heyting structure, a
major traditional objection to multi-valued
logics---that the exact structure of the logic, or
associated algebras, seems arbitrary---does not
apply here.

\end{enumerate}

\subsection{Toposes of Presheaves}
\label{Subsec:TopPre} In preparation for the
applications in Section 3, we turn now to the
theory of presheaves: more precisely, the theory
of presheaves on an arbitrary `small' category
$\cal C$ (the qualification `small' means that the
collection of objects in $\cal C$ is a genuine
set, as is the collection of all $\cal C$'s
arrows).

To make the necessary definition we recall the
idea of a `functor' between a pair of categories
$\cal C$ and $\cal D$. Broadly speaking, this is a
arrow-preserving function from one category to the
other. The precise definition is as follows.

\begin{definition}
\
\begin{itemize}
\item {A {\em covariant functor\/} $\bf F$ from a category
$\cal C$ to a category $\cal D$ is a function that
assigns
    \begin{enumerate}
        \item to each $\cal C$-object $A$, a $\cal D$-object
        ${\bf F}(A)$;

        \item {to each $\cal C$-arrow $f:B\rightarrow A$, a
$\cal D$-arrow ${\bf F}(f):{\bf F}(B)\rightarrow
{\bf F}(A)$ such that ${\bf F}({\rm id}_A)={\rm
id}_{{\bf F}(A)}$; and, if $g:C\rightarrow B$, and
$f:B\rightarrow A$ then
    \begin{equation}
        {\bf F}(f\circ g)={\bf F}(f)\circ
                {\bf F}(g).     \label{Def:covfunct}
    \end{equation}
        }
    \end{enumerate}
    }
\end{itemize}
\end{definition}
A {\em presheaf\/} (also known as a {\em varying
set\/}) on the category $\cal C$ is defined to be
a covariant functor $\bf X$ from the category
$\cal C$ to the category `${\rm Set}$' of normal
sets. We want to make the collection of presheaves
on $\cal C$ into a category, and therefore we need
to define what is meant by an `arrow' between two
presheaves $\bf X$ and $\bf Y$. The intuitive idea
is that such an arrow from $\bf X$ to $\bf Y$ must
give a `picture' of $\bf X$ within $\bf Y$.
Formally, such an arrow is defined to be a {\em
natural transformation\/} $N:{\bf
X}\rightarrow{\bf Y}$, by which is meant a family
of maps (called the {\em components\/} of $N$)
$N_A:{\bf X}(A)\rightarrow{\bf Y}(A)$, $A$ an
object in $\cal C$, such that if $f:A\rightarrow
B$ is an arrow in $\cal C$, then the composite map
${\bf X}(A) \stackrel{N_A}\longrightarrow{\bf
Y}(A)\stackrel{{\bf Y}(f)} \longrightarrow{\bf
Y}(B)$ is equal to ${\bf X}(A) \stackrel{{\bf
X}(f)}\longrightarrow{\bf X}(B)\stackrel{N_B}
\longrightarrow {\bf Y}(B)$. In other words, we
have the commutative diagram
\begin{equation}
    \bundle{{\bf X}(A)}{N_A}{{\bf Y}(A)}
    \bundlemap{{\bf X}(f)}{{\bf Y}(f)}
    \bundle{{\bf X}(B)}{N_B}{{\bf Y}(B)}    \label{cdNT}
\end{equation}
The category of presheaves on $\cal C$ equipped
with these arrows is denoted ${\rm Set}^{\cal C}$.

We say that $\bf K$ is a {\em subobject\/} of $\bf
X$ if there is an arrow in the category of
presheaves ({\em i.e.}, a natural transformation)
$i:{\bf K}\rightarrow{\bf X}$ with the property
that, for each $A$, the component map $i_A:{\bf
K}(A)\rightarrow{\bf X}(A)$ is a subset embedding,
{\em i.e.}, ${\bf K}(A)\subseteq {\bf X}(A)$.
Thus, if $f:A\rightarrow B$ is any arrow in $\cal
C$, we get the analogue of the commutative diagram
Eq.\ (\ref{cdNT}):
\begin{equation}
\bundle{{\bf K}(A)}{}{{\bf X}(A)} \bundlemap{{\bf
K}(f)}{{\bf X}(f)} \bundle{{\bf K}(B)}{}{{\bf
X}(B)} \label{subobject}
\end{equation}
where, once again, the vertical arrows are subset
inclusions.

The category of presheaves on $\cal C$, ${\rm
Set}^{{\cal C}}$, forms a topos. As we have said,
we will not need the full definition of a topos;
but we do need the idea that a topos has a
subobject classifier $\Omega$, to which we now
turn.

\paragraph*{2.3.1 Sieves and the Subobject Classifier in a Topos of Presheaves}

Among the key concepts in presheaf theory---and
something of particular importance for this
paper---is that of a `sieve', which plays a
central role in the construction of the subobject
classifier in the topos of presheaves on a
category $\cal C$.

A {\em sieve\/} on an object $A$ in $\cal C$ is
defined to be a collection $S$ of arrows
$f:A\rightarrow B$ in $\cal C$ with the property
that if $f:A\rightarrow B$ belongs to $S$, and if
$g:B\rightarrow C$ is any arrow, then $g\circ
f:A\rightarrow C$ also belongs to $S$. In the
simple case where $\cal C$ is a poset, a sieve on
$p\in\cal C$ is any subset $S$ of $\cal C$ such
that if $r\in S$ then (i) $p\leq r$, and (ii)
$r'\in S$ for all $r\leq r'$; in other words, a
sieve is nothing but a {\em upper\/} set in the
poset.

The presheaf ${\bf\Omega}:{\cal C}\rightarrow {\rm
Set}$ is now defined as follows. If $A$ is an
object in $\cal C$, then ${\bf\Omega}(A)$ is
defined to be the set of all sieves on $A$; and if
$f:A\rightarrow B$, then
${\bf\Omega}(f):{\bf\Omega}(A)\rightarrow{\bf\Omega}(B)$
is defined as
\begin{equation}
{\bf\Omega}(f)(S):= \{h:B\rightarrow C\mid h\circ
f\in S\}
                                \label{Def:Om(f)}
\end{equation}
for all $S\in{\bf\Omega}(A)$.

For our purposes in what follows, it is important
to note that if $S$ is a sieve on $A$, and if
$f:A\rightarrow B$ belongs to $S$, then from the
defining property of a sieve we have
\begin{equation}
        {\bf\Omega}(f)(S):=\{h:B\rightarrow C\mid h\circ f\in S\}=
\{h:B\rightarrow C\}=:\ \uparrow\!\!B \label{f*S}
\end{equation}
where $\uparrow\!\!B$ denotes the {\em
principal\/} sieve on $B$, defined to be the set
of all arrows in $\cal C$ whose domain is $B$.

If $\cal C$ is a poset, the associated operation
on sieves corresponds to a family of maps
$\Omega_{qp}:\Omega_p\rightarrow\Omega_q$ (where
$\Omega_p$ denotes the set of all sieves on $p$ in
the poset) defined by
$\Omega_{qp}={\bf\Omega}(i_{pq})$ if
$i_{pq}:p\rightarrow q$ ({\em i.e.}, $p\leq q$).
It is straightforward to check that if
$S\in\Omega_q$, then
\begin{equation}
\Omega_{qp}(S):=\uparrow\!{p}\cap S
\label{Def:Omqp}
\end{equation}
where $\uparrow\!{p}:=\{r\in{\cal C}\mid p\leq
r\}$.

A crucial property of sieves is that the set
${\bf\Omega}(A)$ of sieves on $A$ has the
structure of a Heyting algebra. Recall from
Section 1.3.1 that this is defined to be a
distributive lattice, with null and unit elements,
that is relatively complemented---which means that
for any pair $S_1,S_2$ in ${\bf \Omega}(A)$, there
exists an element $S_1\Rightarrow S_2$ of
${\bf\Omega}(A)$ with the property that, for all
$S\in{\bf\Omega}(A)$,
\begin{equation}
        S\leq (S_1\Rightarrow S_2)\mbox{  if and only if
$S\land S_1\leq S_2$}.
\end{equation}
Specifically, ${\bf\Omega}(A)$ is a Heyting
algebra where the unit element
$1_{{\bf\Omega}(A)}$ in ${\bf\Omega}(A)$ is the
principal sieve $\uparrow\!\!A$, and the null
element $0_{{\bf\Omega}(A)}$ is the empty sieve
$\emptyset$. The partial ordering in
${\bf\Omega}(A)$ is defined by $S_1\leq S_2$ if,
and only if, $S_1\subseteq S_2$; and the logical
connectives are defined as:
\begin{eqnarray}
    && S_1\land S_2:=S_1\cap S_2    \label{Def:S1landS2}\\
    && S_1\lor S_2:=S_1\cup S_2     \label{Def:S1lorS2} \\
    &&S_1\Rightarrow S_2:=\{f:A\rightarrow B\mid
    \mbox{ for all $g:B\rightarrow C$ if $g\circ f\in S_1$ then
                $g\circ f\in S_2$}\}.
\end{eqnarray}
As in any Heyting algebra, the negation of an
element $S$ (called the {\em pseudo-complement\/}
of $S$) is defined as $\neg S:=S\Rightarrow 0$; so
that
\begin{equation}
    \neg S:=\{f:A\rightarrow B\mid \mbox{for all
$g:B\rightarrow C$, $g\circ f\not\in S$} \}.
\label{Def:negS}
\end{equation}
The main distinction between a Heyting algebra and
a Boolean algebra is that, in the former, the
negation operation does not necessarily obey the
law of excluded middle: instead, all that be can
said is that, for any element $S$,
\begin{equation}
        S\lor\neg S\leq 1.
\end{equation}

It can be shown that the presheaf ${\bf\Omega}$ is
a subobject classifier for the topos ${\rm
Set}^{{\cal C}}$. That is to say, subobjects of
any object $\bf X$ in this topos ({\em i.e.}, any
presheaf on $\cal C$) are in one-to-one
correspondence with arrows $\chi:{\bf
X}\rightarrow {\bf\Omega}$. This works as follows.
First, let $\bf K$ be a subobject of $\bf X$. Then
there is an associated {\em characteristic\/}
arrow $\chi^{{\bf K}}:{\bf
X}\rightarrow{\bf\Omega}$, whose `component'
$\chi^{{\bf K}}_A:{\bf
X}(A)\rightarrow{\bf\Omega}(A)$ at each `stage of
truth' $A$ in $\cal C$ is defined as
\begin{equation}
    \chi^{{\bf K}}_A(x):=\{f:A\rightarrow B\mid {\bf X}(f)(x)\in
{\bf K}(B)\} \label{Def:chiKA}
\end{equation}
for all $x\in {\bf X}(A)$. That the right hand
side of Eq.\ (\ref{Def:chiKA}) actually {\em is\/}
a sieve on $A$ follows from the defining
properties of a subobject.

Thus, in each `branch' of the category $\cal C$
going `upstream' from the stage $A$, $\chi^{{\bf
K}}_A(x)$ picks out the first member $B$ in that
branch for which ${\bf X}(f)(x)$ lies in the
subset ${\bf K}(B)$, and the commutative diagram
Eq.\ (\ref{subobject}) then guarantees that ${\bf
X}(h\circ f)(x)$ will lie in ${\bf K}(C)$ for all
$h:B\rightarrow C$. Thus each `stage of truth' $A$
in $\cal C$ serves as a possible context for an
assignment to each $x\in {\bf X}(A)$ of a
generalised truth-value: which is a sieve,
belonging to the Heyting algebra ${\bf\Omega}(A)$,
rather than an element of the Boolean algebra
$\{0,1\}$ of normal set theory. This is the sense
in which contextual, generalised truth-values
arise naturally in a topos of presheaves.

There is a converse to Eq.\ (\ref{Def:chiKA}):
namely, each arrow $\chi:{\bf
X}\rightarrow{\bf\Omega}$ ({\em i.e.}, a natural
transformation between the presheaves ${\bf X}$
and ${\bf\Omega}$) defines a subobject ${\bf
K}^\chi$ of $\bf X$ via
\begin{equation}
    {\bf K}^\chi(A):=\chi_A^{-1}\{1_{{\bf\Omega}(A)}\}.
                            \label{Def:KchiA}
\end{equation}
at each stage of truth $A$.

\paragraph*{2.3.2 Global Sections of a Presheaf}

For the category of presheaves on $\cal C$, a
terminal object ${\bf 1}:{\cal C}\rightarrow {\rm
Set}$ can be defined by ${\bf 1}(A):=\{*\}$ at all
stages $A$ in $\cal C$; if $f:A\rightarrow B$ is
an arrow in $\cal C$ then ${\bf
1}(f):\{*\}\rightarrow\{*\}$ is defined to be the
map $*\mapsto *$. This is indeed a terminal object
since, for any presheaf $\bf X$, we can define a
unique natural transformation $N:{\bf
X}\rightarrow{\bf 1}$ whose components $N_A:{\bf
X}(A)\rightarrow{\bf 1}(A)=\{*\}$ are the constant
maps $x\mapsto *$ for all $x\in{\bf X}(A)$.

A global element (or point) of a presheaf $\bf X$
is also called a {\em global section\/}. As an
arrow $\gamma:{\bf 1}\rightarrow{\bf X}$ in the
topos ${\rm Set}^{{\cal C}}$, a global section
corresponds to a choice of an element
$\gamma_A\in{\bf X}(A)$ for each stage of truth
$A$ in $\cal C$, such that, if $f:A\rightarrow B$,
the `matching condition'
\begin{equation}
    {\bf X}(f)(\gamma_A)=\gamma_B \label{Def:global}
\end{equation}
is satisfied. As we shall see, the Kochen-Specher
theorem can be read as asserting the non-existence
of any global sections of certain presheaves that
arises naturally in any quantum theory.

\section{Some Presheaves in Quantum Theory and Quantum Gravity}
\label{Sec:SomPreQTQG} Having developed in Section
2 the idea of a topos, especially the idea of a
topos of presheaves, we wish now to suggest some
possible applications in quantum physics.

There are several natural orders in which to
present these examples. For instance, one could
follow Section 1's order of first treating quantum
theory, without regard to space, time or gravity;
and then treating these latter. But we will in
fact proceed by first giving several examples
involving space, time or spacetime, since: (i) in
these examples, it is especially natural to think
of the objects of the presheaf's base-category
$\cal C$ as `contexts' or `stages' relative to
which generalized truth-values are assigned; and
(ii) these examples will serve as prototypes, in
various ways, for later examples.

\paragraph{3.1 Global reference frames in elementary wave mechanics}
\label{Para:GloCooSysEleWavMec} Throughout
classical and quantum physics, we are often
concerned with reference frames (or coordinate
systems), the transformations between them, and
the corresponding transformations on states of a
physical system, and on physical quantities. Our
first example will present in terms of presheaves
some familiar material about reference frames in
the context of non-relativistic wave mechanics.

Define the category of contexts $\cal C$ to have
as its objects global Cartesian reference frames
$e:=\{e^1,e^2,e^3\}$ (where $e^i$, $i=1,2,3$, are
vectors in Euclidean 3-space $E^3$ such that
$e^i\cdot e^j=\delta^{ij}$), all sharing a common
origin; and define $\cal C$ to have as its arrows
the orthogonal transformations $O(e,e')$ from one
reference frame $\{e^i\}$ to another $\{e'^i\}$,
{\em i.e.}, with a matrix representation $e'^i =
\sum_{j=1}^3e^j O(e,e')_j^i$; (so that between any
two objects, there is a unique arrow). Define a
presheaf $\bf H$ as assigning to each object $e$
in $\cal C$, a copy ${\bf H}(e)$ of the Hilbert
space $L^2(\mathR^3)$; and to each arrow
$O(e,e')$, the unitary map $U(e,e'): {\bf H}(e)
\rightarrow {\bf H}(e')$ defined by
$(U(e,e')\psi)(x) : = \psi(O(e,e')^{-1}(x))$ (so
that $U(e,e')$ represents the action of $O(e,e')$
as a map from one copy, ${\bf H}(e)$, of the
(pure) state-space $L^2(\mathR^3)$, to the other
copy ${\bf H}(e')$). Any given $\psi\in
L^2(\mathR^3)$, together with its transforms under
the various unitary maps $U(e,e')$, defines a
global section of $\bf H$.

Of course, discussions of the transformation of
the wave-function under spatial rotations etc.\
normally identify the different copies of the
state-space $L^2(\mathR^3)$; and from the
viewpoint of those discussions, the above
definition of $\bf H$ may seem at first sight to
make a mountain out of a molehill, particularly
since the category of contexts in this example is
so trivial (for example, the internal logic is
just the standard `true-false' logic). But it is a
helpful prototype to have in mind when we come to
more complex or subtle examples.

For the moment, just note that this definition has
the advantage of clearly distinguishing the
quantum state at the given time from its
representing vectors $\psi$ in various reference
frames. Or rather, to be precise, we need to allow
for the fact that the quantum state is a yet more
abstract notion, also occurring in other
representations than wave mechanics
(position-representation). So the point is: this
definition of $\bf H$ distinguishes the
Schr\"odinger-picture, wave-mechanical
representative of the quantum state at the given
time---which it takes as a global section of $\bf
H$---from its representing vectors $\psi$
(elements of the global section at the various
`stages' $e$).

\paragraph{3.2 Observers in quantum cosmology}
One could argue that the example above illustrates
a contextual aspect of standard quantum theory
whereby the concrete representation of an abstract
state depends on the observer; at least, this is
so if we identify reference frames with observers.
As we have explained, this contextual aspect is
not emphasised in standard quantum theory since
the different Hilbert spaces associated with
different observers are all naturally isomorphic
(via the unitary operators $U(e,e'):{\bf
H}(e)\rightarrow{\bf H}(e')$). From a physical
perspective, the fact that different observers,
related by a translation or a rotation of
reference frame, see `equivalent' physics is a
reflection of the homogeneity and isotropy of
physical space.

However, the situation might well be different in
cosmological situations, since the existence of
phenomena like event and particle horizons means
that the physics perceptible from the perspective
of one observer may be genuinely different from
that seen by another. This suggests that any
theory of quantum cosmology (or even quantum field
theory in a fixed cosmological background) may
require the use of more than one Hilbert space, in
a way that cannot be `reduced' to a single space.

Of course, it is well known that quantum field
theory on a curved spacetime often requires more
than one Hilbert space, associated with the
unavoidable occurrence of inequivalent
representations of the canonical commutation
relations: this is one of the reasons for
preferring a $C^*$-algebra approach. But what we
have in mind is different---for example, our
scheme could easily be adapted to involve a
presheaf of $C^*$-algebras, each associated with
an `observer'.

Evidently, a key question in this context is what
is meant by an `observer'; or, more precisely, how
this idea should be represented mathematically in
the formalism. One natural choice might be a
time-like curve (in the case of quantum field
theory in a curved background with horizons),
although this does suggest that a `history'
approach to quantum theory would be more
appropriate than any of the standard ones. Of
course, in the case of quantum cosmology proper,
these issues become far more complex since---for
example---even what is meant by a `time-like
curve' presumably becomes the subject of quantum
fluctuations!

\paragraph{3.3 Unitary time evolution in elementary quantum theory}
\label{Para:TimEvoEleQuaThe} The discussion above
of different spatial reference frames has a
precise temporal analogue. Thus, fix once for all
a global Cartesian reference frame in $E^3$, and
define the base-category of contexts $\cal C$ to
be the real line $\mathR$, representing time. That
is to say, let the objects of $\cal C$ be instants
$t\in \mathR$; and let there be an $\cal C$-arrow
from $t$ to $t'$, $f:t \rightarrow t'$, if and
only if $t \leq t'$; so there is at most one arrow
between any pair of objects $t, t'$ in $\cal C$.
Define the presheaf, called $\bf H$ (as in
Paragraph 3.1), as assigning to each $t$ a copy of
the system's Hilbert space $\cal H$; ($\cal H$
need not be $L^2(\mathR^3)$---here we generalize
from wave mechanics). Writing this copy as ${\cal
H}_t$, we have ${\bf H}(t) := {\cal H}_t$. The
action of $\bf H$ on ${\cal C}$-arrows is defined
by the Hamiltonian $\hat H$, via its one-parameter
family of unitary exponentiations $U_t$. If $f:t
\rightarrow t'$, then ${\bf H}(f):{\cal H}_t
\rightarrow {\cal H}_{t'}$ is defined by $U_{t' -
t}$. The action of $U_{t' - t}$, then represents
the Schr\"odinger-picture evolution of the system
from time $t$ to $t'$; and a total history of the
system (as described in the given spatial
coordinate system) is represented by a global
section of the presheaf $\bf H$.\footnote{We
remark that we could similarly express in terms of
presheaves Heisenberg-picture evolution: we would
instead define a presheaf that assigned to each
$\cal C$-object $t$ a copy of the set $B({\cal
H})$ of bounded self-adjoint operators on $\cal H$
(or say, a copy of some other fixed set taken as
the algebra of observables), and then have the
maps $U_t$ induce Heisenberg-picture evolution on
the elements of the copies of $B({\cal H})$.} Note
that, as in the example of 3.1, the internal logic
of this example is essentially trivial.

 We note that a parallel discussion could be given for
time evolution in classical physics: we would
attach a copy of the phase-space $\Gamma$ to each
$t$, and a total history of the system (as
described in the given spatial coordinate system)
would be represented by a global section of the
corresponding presheaf. It transpires that the
development of such a `history' approach to
classical physics provides a very illuminating
perspective on the mathematical structures used in
the consistent-histories approach to quantum
theory; for more information see \cite{Sav98}.

\noindent This completes the presentation, in
terms of presheaves, of familiar material from
orthodox/established theories. From now on, we
will present in terms of presheaves some ideas
that are currently being pursued in research on
foundations of quantum theory and quantum gravity.

\paragraph{3.4 Presheaves on causal sets}
\label{Para:PrePoiRegCauSet} The previous example
admits an immediate generalisation to the theory
of causal sets. By a {\em causal set\/} we mean a
partially-ordered set ${\cal P}$ whose elements
represent spacetime points in a discrete,
non-continuum model, and in which $p\leq q$, with
$p,q\in{\cal P}$, means that $q$ lies in the
causal future of $p$.

The set $\cal P$ is a natural base category for a
presheaf of Hilbert spaces in which the Hilbert
space at a point $p\in\cal P$ represents the
quantum degrees of freedom that are `localised' at
that point/context. From another point of view,
the Hilbert space at a point $p$ could represent
the history of the system (thought of now in a
cosmological sense) as viewed from the perspective
of an observer localised at that point. For a
discussion of this idea see \cite{Mar98}. The
sieve, and hence logical, structure in this
example is distinctly non-trivial.

There are several variants on this theme: for
example, one may decide that the category of
contexts should have as its objects `regions'
rather than spacetime points (cf.\ Section 1.4.1
in this regard).

\paragraph{3.5 Presheaves on spatial slices; for topological QFT}
\label{Para:PreSpaSliTopQFT}

Topological quantum field theory (TQFT) has a very
well-known formulation in terms of category
theory, and it is rather straightforward to see
that this extends naturally to give a certain
topos perspective.

 Recall that in differential topology, two closed
$n$-dimensional manifolds $\Sigma_1$ and
$\Sigma_2$ are said to be {\em cobordant\/} if
there is a compact $n+1$-manifold, $M$ say, whose
boundary $\partial M$ is the disjoint union of
$\Sigma_1$ and $\Sigma_2$. In TQFT, the
$n$-dimensional manifolds are interpreted as
possible models for physical space (so that
spacetime has dimension $n+1$), and an
interpolating $n+1$-manifold is thought of as
describing a form of `topology change' in the
context of a (euclideanised) type of quantum
gravity theory. In the famous Atiyah axioms for
TQFT, a Hilbert space ${\cal H}_\Sigma$ is
attached to each spatial $n$-manifold $\Sigma$,
and to each cobordism from $\Sigma_1$ to
$\Sigma_2$ there is associated a unitary map from
${\cal H}_{\Sigma_1}$ to ${\cal H}_{\Sigma_2}$.

We note that the collection of all compact
$n$-dimensional manifolds can be regarded as the
set of objects in a category $\cal C$, in which
the arrows from an object $\Sigma_1$ to another
$\Sigma_2$ are given by cobordisms from $\Sigma_1$
to $\Sigma_2$. From this perspective, the Atiyah
axioms for TQFT can be viewed as a statement of
the existence of a functor from $\cal C$ to the
category of Hilbert spaces; indeed, this is how
these axioms are usually stated. However, from the
perspective being developed in the present paper,
we see that we can also think of $\cal C$ as a
`category of contexts', in which case we have a
natural presheaf reformulation of TQFT.

\paragraph{3.6 Consistent histories formalism for quantum theory and continuous time}
\label{Para:TwoTimConHisQuaThe}

In the `History Projection Operator' (HPO) version
of the consistent-histories approach to quantum
theory, propositions about the history of the
system at a finite set of time points
$(t_1,t_2,\ldots,t_n)$ are represented by
projection operators on the tensor product ${\cal
H}_{t_1}\otimes{\cal H}_{t_2}\otimes
\cdots\otimes{\cal H}_{t_n}$ of $n$ copies of the
Hilbert space $\cal H$ associated with the system
by standard quantum theory. The choice of this
particular Hilbert space can be motivated in
several different ways. For example, the original
motivation \cite{I94} was a desire to find a
concrete representation of the temporal logic of
such history propositions. From another
perspective, this Hilbert space can be seen as the
carrier of an irreducible representation of the
`history group' whose Lie algebra is (on the
simplifying assumption that the system is a
non-relativistic point particle moving in one
dimension)
\begin{eqnarray}
[x_{t_i},x_{t_j}]&=&0 \label{HAxx} \\
{[}p_{t_i},p_{t_j}]&=&0 \label{HApp} \\
{[}x_{t_i},p_{t_j}]&=&i\hbar\delta_{ij}\label{HAxp}
\end{eqnarray}
where $i,j=1,2,\ldots, n$, and $x_{t_i}$ (resp.
$p_{t_i}$) is the Schr\"odinger-picture operator
whose spectral projectors represent propositions
about the position (resp. momentum) of the system
at the time $t_i$.

One advantage of the approach based on equations
(\ref{HAxx}--\ref{HAxp}) is that it suggests an
immediate generalisation to the case of {\em
continuous\/}-time histories: namely, the use of
the history algebra
\begin{eqnarray}
[x_{t},x_{t'}]&=&0 \label{CTHAxx} \\
{[}p_{t},p_{t'}]&=&0 \label{CTHApp} \\
{[}x_{t},p_{t'}]&=&i\hbar\tau\delta(t'-t)
\label{CTHAxp}
\end{eqnarray}
where $\tau$ is a constant with the dimensions of
time.

This continuous-time history algebra has been
studied by a variety of authors but here we will
concentrate on Savvidou's observation \cite{Sav98}
that the notion of `time' appears in two ways that
differ in certain significant respects. The main
idea is to introduce a new time coordinate $s \in
\mathR$, and to associate with it a Heisenberg
picture defined from the time-averaged Hamiltonian
$H=\int dt H_t$. Thus, in particular, one defines
for the time-indexed position operator $x_t$
\begin{equation}
x_{t}(s) := \exp(isH/\hbar)\,x_t\,\exp(-isH/\hbar)
\end{equation}
This new time $s$ is {\em not\/} a difference in
values of $t$. Rather, if one thinks of assigning
a copy ${\cal H}_t$ of the system's (usual)
Hilbert space $\cal H$ to each time $t$, then $s$
parametrizes a Heisenberg-picture motion of
quantities {\em within\/} ${\cal H}_t$.
Accordingly, $t$ is called `external time', and
$s$ is called `internal time'.

This formalism has been developed in various ways:
in particular, there is a natural, dynamics-{\em
independent\/} `Liouville' operator that generates
translations in the external time parameter. From
our topos-theoretic perspective, we note that
external time is more singular than internal
time---as hinted by the delta-functions in $t$
that occur in the history algebra's canonical
commutation relations. This suggests modelling
external time, not by the usual real numbers
$\mathR$, but by the reals `enriched' with
infinitesimals in the sense of synthetic
differential geometry, and which are related in
some way to the action of the Liouville operator.
As emphasised earlier, this requires a
non-standard model of the real line: in fact, we
have to use a real number object in a topos.

Note that this use of a topos is quite different
from, and in addition to, any development of a
consistent-histories analogue of the temporal
presheaf introduced in Section 3.3. In the latter
case, the presheaf structure in the
consistent-histories theory can arguably be
related to ideas of state reduction of the kind
discussed by von Neumann and L\"uders
\cite{Sav98}.

\paragraph{3.7 Presheaves of Propositions, and Valuations in Quantum Theory}
\label{Para:PreProValQuaThe} Finally, we want to
briefly present an application of presheaves that
we have developed in detail elsewhere
\cite{IB98,IB99,HIB99}. Namely: the proposal
mentioned at the end of Section 1.1, to retain a
`realist flavour' in the assignment of values to
quantum-theoretic quantities, despite no-go
theorems like the Kochen-Specker theorem, by using
the non-Boolean logical structure of a topos of
presheaves.

Before stating the proposal precisely, let us
motivate it in terms the assumptions (i) and (ii)
of Section 1.1. As discussed there, in quantum
theory, assumption (i), {\em i.e.}, that all
quantities have real-number values, fails by
virtue of the Kochen-Specker theorem; and
assumption (ii), that one can measure any quantity
ideally, is very problematic, involving as it does
the notion of measurement. Standard quantum
theory, with its `eigenvalue-eigenstate
link'---that in state $\psi$ there is a value only
for a quantity of which $\psi$ is an eigenstate,
viz. the eigenvalue---retains assumption (ii) only
in the very limited sense that {\em if} the
quantity $A$ has a value, $r$ say, according to
the theory, {\em i.e.}, the (pure) state $\psi$ is
an eigenvector of $\hat{A}$ for eigenvalue $r$,
then an ideal measurement of $A$ would have result
$r$. But setting aside this very special case, the
theory faces the notorious `measurement problem':
the scarcity of values in the microrealm, due to
the eigenvalue-eigenstate link, threatens to make
the macrorealm indefinite (`Schr\"odinger's cat').

 This is of course not the place to review the programmes for
solving this problem. But it is worth
distinguishing two broad approaches to it, which
we will call `Literalism' and `Extra
Values'.\footnote{We give a more detailed
discussion of these approaches and two others,
especially with a view to quantum gravity, in
Section 2.1 of \cite{BI99}.} For our
topos-theoretic proposal will combine aspects of
these approaches. They are:
\begin{enumerate}
\item Literalism. This approach
aims to avoid the instrumentalism of standard
quantum theory, and yet retain its scarcity of
values (the eigenvalue-eigenstate link), while
solving the measurement problem: not by
postulating a non-unitary dynamics, but by a
distinctively {\em interpretative\/} strategy. So
far as we know, there are two main forms of this
approach: Everettian views (where the
eigenvalue-eigenstate link is maintained `within a
branch'); and those based on quantum logic.

\item Extra Values. This approach gives up the
eigenvalue-eigenstate link; but retains standard
quantum theory's unitary dynamics for the quantum
state. It postulates extra values (and equations
for their time-evolution) for some quantities. The
quantities getting these extra values are selected
either {\em a priori\/}, as in the pilot-wave
programme, or by the quantum state itself, as in
(most) modal interpretations.
\end{enumerate}

Our topos-theoretic proposal combines aspects of
Literalism and Extra Values. Like both these
approaches, the proposal is `realist', not
instrumentalist; (though it also shares with
standard quantum theory, at least in its Bohrian
or `Copenhagen' version, an emphasis on
contextuality). Like Extra Values (but unlike
Literalism), it attributes values to quantities
beyond those ascribed by the eigenvalue-eigenstate
link. Like Literalism (but unlike Extra Values),
these additional values are naturally defined by
the orthodox quantum formalism. More specifically:
{\em all\/} quantities get additional values (so
no quantity is somehow `selected' to get such
values); any quantum state defines such a
valuation, and any such valuation obeys an
appropriate version of the {\em FUNC\/} constraint
mentioned in Section 1.1. The `trick', whereby
such valuations avoid no-go theorems like the
Kochen-Specker theorem, is that the truth value
ascribed to a proposition about the value of a
physical quantity is not just `true' or `false'!

Thus consider the proposition ``$A \in \Delta$'',
saying that the value of the quantity $A$ lies in
a Borel set $\Delta \subseteq \mathR $. Roughly
speaking, any such proposition is ascribed as a
truth-value a set of coarse-grainings,
$f(\hat{A})$, of the operator $\hat{A}$ that
represents $A$. Exactly which coarse-grainings are
in the truth-value depends in a precise and
natural way on $\Delta$ and the quantum state
$\psi$: in short, $f(\hat{A})$ is in the
truth-value iff $\psi$ is in the range of the
spectral projector $\hat{E}[f(A) \in f(\Delta)]$.
Note the contrast with the eigenstate-eigenvalue
link: our requirement is not that $\psi$ be in the
range of $\hat{E}[A \in \Delta]$, but a weaker
requirement. For $\hat{E}[f(A) \in f(\Delta)]$ is
a larger spectral projector; {\em i.e.\/}, in the
lattice $\cal L(H)$ of projectors on the Hilbert
space $\cal H$, $\hat{E}[A \in \Delta] <
\hat{E}[f(A) \in f(\Delta)]$.

 So the intuitive idea is that the new proposed truth-value of ``$A
\in \Delta$'' is given by the set of weaker
propositions ``$f(A) \in f(\Delta)$'' that are
true in the old ({\em i.e.}, eigenstate-eigenvalue
link) sense. To put it a bit more exactly: the new
proposed truth-value of ``$A \in \Delta$'' is
given by the set of quantities $f(A)$ for which
the corresponding weaker proposition ``$f(A) \in
f(\Delta)$'' is true in the old ({\em i.e.},
eigenstate-eigenvalue link) sense. To put it {\em
less\/} exactly, but more memorably: the new
truth-value of a proposition is given by the set
of its consequences that are true in the old
sense.

We turn to stating the proposal exactly. We first
introduce the set ${\cal O}$ of all bounded
self-adjoint operators $\hat A,\hat B,\ldots$ on
the Hilbert space $\cal H$ of a quantum system. We
turn ${\cal O}$ into a category by defining the
objects to be the elements of ${\cal O}$, and
saying that there is an arrow from $\hat A$ to
$\hat B$ if there exists a real-valued function
$f$ on $\sigma(\hat A)\subset\mathR$, the spectrum
of $\hat A$, such that $\hat B=f(\hat A)$ (with
the usual definition of a function of a
self-adjoint operator, using the spectral
representation). If $\hat B=f(\hat A)$, for some
$f:\sigma(\hat A)\rightarrow\mathR$, then the
corresponding arrow in the category ${\cal O}$
will be denoted $f_{{\cal O}}: \hat
A\rightarrow\hat B$.

We next define two presheaves on the category
${\cal O}$, called the {\em dual presheaf\/} and
the {\em coarse-graining presheaf\/} respectively.
The former affords an elegant formulation of the
Kochen-Specker theorem, namely as a statement that
the dual presheaf does not have global sections.
The latter is at the basis of our proposed
generalised truth-value assignments.

The dual presheaf on ${{\cal O}}$ is the covariant
functor ${\bf D}:{\cal O}\rightarrow {\rm Set}$
defined as follows:
\begin{enumerate}
    \item On objects: ${\bf D}(\hat A)$ is the {\em dual\/} of
$W_A$, where $W_A$ is the spectral algebra of the
operator $\hat A$; {\em i.e.} $W_A$ is the
collection of all projectors onto the subspaces of
$\cal H$ associated with Borel subsets of
$\sigma(\hat A)$. That is to say: ${\bf D}(\hat
A)$ is defined to be the set ${\rm
Hom}(W_A,\{0,1\})$ of all homomorphisms from the
Boolean algebra $W_A$ to the Boolean algebra
$\{0,1\}$.

    \item On arrows: If $f_{{\cal O}}:\hat A\rightarrow \hat B$,
so that $\hat B=f(\hat A)$, then ${\bf D}(f_{\cal
O}):D(W_A)\rightarrow D(W_B)$ is defined by ${\bf
D}(f_{\cal O})(\chi):= \chi|_{W_{f(A)}}$ where
$\chi|_{W_{f(A)}}$ denotes the restriction of
$\chi\in D(W_A)$ to the subalgebra
$W_{f(A)}\subseteq W_A$.

\end{enumerate}

    A global element (global section) of the functor ${\bf
D}:{\cal O}\rightarrow {\rm Set}$ is then a
function $\gamma$ that associates to each ${\hat
A} \in\cal O$ an element $\gamma_{A}$ of the dual
of $W_A$ such that if $f_{\cal O}:{\hat
A}\rightarrow {\hat B}$ (so $\hat B=f(\hat A)$ and
$W_{B}\subseteq W_A$), then
$\gamma_{A}|_{W_B}=\gamma_{B}$. Thus, for all
projectors $\hat\alpha\in W_{B}\subseteq W_A$,
\begin{equation}
\gamma_{B}(\hat\alpha)=\gamma_{A}(\hat\alpha).
\label{Funcgamma}
\end{equation}

Since each $\hat\alpha$ in the lattice $\cal
{L(H)}$ of projection operators on $\cal H$
belongs to at least one such spectral algebra
$W_A$ (for example, the algebra $\{\hat 0,\hat
1,\hat\alpha,\hat 1 - \hat\alpha\}$) it follows
from Eq. (\ref{Funcgamma}) that a global section
of ${\bf D}$ associates to each projection
operator $\hat\alpha \in \cal {L(H)}$ a number
$V(\hat\alpha)$ which is either $0$ or $1$, and is
such that, if $\hat\alpha\land\hat\beta=\hat 0$,
then
$V(\hat\alpha\lor\hat\beta)=V(\hat\alpha)+V(\hat\beta)$.
In other words, a global section $\gamma$ of the
presheaf ${\bf D}$ would correspond to an
assignment of truth-values $\{0,1\}$ to all
propositions of the form ``$A\in\Delta$'', which
obeyed the {\em FUNC} condition Eq.\
(\ref{Funcgamma}). These are precisely the types
of valuation prohibited, provided that $\dim{\cal
H}>2$, by the Kochen-Specker theorem. So an
alternative way of expressing the Kochen-Specker
theorem is that, if $\dim{\cal H}>2$, the dual
presheaf ${\bf D}$ has no global sections.

However, we {\em can} use the subobject classifier
$\bf \Omega$ in the topos ${\rm Set}^{\cal O}$ of
all presheaves on $\cal O$ to assign {\em
generalized} truth-values to the propositions
``$A\in\Delta$''. These truth-values will be
sieves, as defined in Section 2.2; and since they
will be assigned relative to each `context' or
`stage of truth' $\hat A$ in $\cal O$, these
truth-values will be contextual as well as
generalized. And as we said at the end of Section
2.2: because in any topos the subobject classifier
$\bf \Omega$ is fixed by the structure of the
topos, $\bf \Omega$ is unique up to isomorphism.
Thus the family of associated truth-value
assignments is fixed, and the traditional
objection to multi-valued logics---that their
structure often seems arbitrary---does not apply
to these generalized, contextual truth-values.

We first define the appropriate presheaf of
propositions. The {\em coarse-graining presheaf\/}
over $\cal O$ is the covariant functor ${\bf
G}:{\cal O}\rightarrow{\rm Set}$ defined as
follows.
\begin{enumerate}
\item {\em On objects in $\cal O$:} ${\bf G}(\hat A):=
W_A$, where $W_A$ is the spectral algebra of $\hat
A$.

\item {\em On arrows in $\cal O$:} If $f_{\cal O}:\hat A
\rightarrow\hat B$ ({\em i.e.}, $\hat B=f(\hat
A)$), then ${\bf G}(f_{\cal O}): W_A\rightarrow
W_B$ is defined as
\begin{equation}
        {\bf G}(f_{\cal O})(\hat E[A\in\Delta]):=
        \hat E[f(A)\in f(\Delta)]       \label{Def:G(fO)}
\end{equation}
(where, if $f(\Delta)$ is not Borel, the right
hand side is to be understood in the sense of
Theorem 4.1 of \cite{IB98}---a measure-theoretic
nicety that we shall not discuss here).
\end{enumerate}

We call a function $\nu$ that assigns to each
choice of object $\hat A$ in $\cal O$ and each
Borel set $\Delta \subseteq \sigma(\hat A)$, a
sieve of arrows in $\cal O$ on $\hat A$ ({\em
i.e.}, a sieve of arrows with $\hat A$ as domain),
a {\em sieve-valued valuation\/} on $\bf G$. We
write the values of this function as $\nu(A \in
\Delta)$. (One could equally well write $\nu(\hat
E[A\in\Delta])$, provided one bears in mind that
the value depends not only on the projector $\hat
E[A\in\Delta]$, but also on the operator (context)
$\hat A$ of whose spectral family the projector is
considered to be a member.)

From the logical point of view, a natural
desideratum for any kind of valuation on a
presheaf of propositions such as $\bf G$ is that
the valuation should specify a subobject of $\bf
G$. For in logic one often thinks of a valuation
as specifying the `selected' or `winning'
propositions: in this case, the `selected'
elements $\hat E[A\in\Delta]$ in each ${\bf
G}(\hat A)$. So it is natural to require that the
elements that a valuation `selects' at the various
contexts $\hat A$ together define a subobject of
$\bf G$. But we saw in Section 2.3.1 that
subobjects are in one-one correspondence with
arrows, {\em i.e.}, natural transformations,
$N:{\bf G}\rightarrow {\bf\Omega}$. So it is
natural to require a sieve-valued valuation $\nu$
to define such a natural transformation by the
equation $N^\nu_A(\hat E[A\in\Delta]):=\nu(A \in
\Delta)$.

This desideratum leads directly to the analogue
for presheaves of the famous functional
composition condition of the Kochen-Specker
theorem, called {\em FUNC\/} above: and which we
will again call {\em FUNC\/} in the setting of
presheaves. For it turns out that a sieve-valued
valuation defines such a natural transformation
iff it obeys (the presheaf version of) {\em
FUNC\/}.

To spell this out, we first recall that the
subobject classifer $\bf \Omega$ `pushes along'
sieves, according to Eq.\ (\ref{Def:Om(f)}). For
the category $\cal O$, this becomes: if $f_{\cal
O}:\hat A\rightarrow\hat B$, then
${\bf\Omega}(f_{\cal O}):{\bf\Omega}(\hat
A)\rightarrow{\bf\Omega}(\hat B)$ is defined by
\begin{equation}
{\bf\Omega}(f_{\cal O})(S):= \{h_{\cal
O}:B\rightarrow C\mid h_{\cal O}\circ f_{\cal
O}\in S\}
                                \label{Def:Om(f)forO}
\end{equation}
for all sieves $S\in{\bf\Omega}(\hat A)$.

Accordingly, we say that a sieve-valued valuation
$\nu$ on $\bf G$ satisfies {\em generalized
functional composition\/}---for short, {\em
FUNC\/}---if for all $\hat A,\hat B$ and $f_{\cal
O}:\hat A\rightarrow\hat B$ and all ${\hat E}[A
\in \Delta]\in{\bf G}(\hat A)$, the valuation
obeys
\begin{equation}
    \nu(B \in {\bf G}(f)({\hat E}[A \in \Delta])) \equiv \nu(f(A)
\in f(\Delta)) = {\bf\Omega}(f_{\cal O})(\nu(A \in
\Delta)). \label{Def:GenlzdFunc}
\end{equation}

{\em FUNC\/} is exactly the condition a
sieve-valued valuation must obey in order to thus
define a natural transformation, {\em i.e.}, a
subobject of $\bf G$, by the natural equation
$N^\nu_A(\hat E[A\in\Delta]):=\nu(A \in \Delta)$.
That is: A sieve-valued valuation $\nu$ on $\bf G$
obeys {\em FUNC\/} if and only if the functions at
each `stage of truth' $\hat A$
\begin{equation}
    N^\nu_{\hat A}(\hat E[A\in\Delta]):=\nu(A \in \Delta)
\end{equation}
define a natural transformation $N^\nu$ from ${\bf
G}$ to ${\bf\Omega}$.

It turns out that with any quantum state there is
associated such a {\em FUNC\/}-obeying
sieve-valued valuation. Furthermore, this
valuation gives the natural generalization of the
eigenvalue-eigenstate link described at the start
of this Subsection. That is, a quantum state
$\psi$ induces a sieve on each $\hat A$ in $\cal
O$ by the requirement that an arrow $f_{\cal
O}:\hat A\rightarrow\hat B$ is in the sieve iff
$\psi$ is in the range of the spectral projector
$\hat{E}[B \in f(\Delta)]$. To be precise, we
define for any $\psi$, and any $\Delta$ a Borel
subset of the spectrum $\sigma(\hat A)$ of $\hat
A$:
\begin{eqnarray}
    \nu^\psi(A\in\Delta)&:=&\{f_{\cal O}:\hat A\rightarrow\hat B
        \mid \hat E[B\in f(\Delta)]\psi=\psi\} \nonumber\\[2pt]
    &=&  \{f_{\cal O}:\hat A\rightarrow\hat B
        \mid {\rm Prob}(B\in f(\Delta);\psi)=1\}
                \label{Def:nupsiDelta}
\end{eqnarray}
where ${\rm Prob}(B\in f(\Delta);\psi)$ is the
usual Born-rule probability that the result of a
measurement of $B$ will lie in $f(\Delta)$, given
the state $\psi$.

This definition generalizes the
eigenstate-eigenvalue link, in the sense that we
require not that $\psi$ be in the range of
$\hat{E}[A \in \Delta]$, but only that it be in
the range of the larger projector $\hat{E}[f(A)
\in f(\Delta)]$. One can check that the definition
satisfies {\em FUNC\/}, and also has other
properties that it is natural to require of a
valuation (discussed in \cite{IB98,IB99,HIB99}).

 Finally, we note that furthermore, {\em FUNC\/} and these other
properties are enjoyed by the exactly analogous
definition of a sieve-valued valuation $\nu^\rho$
associated with a density matrix $\rho$. One
defines:
\begin{eqnarray}
    \nu^\rho(A\in\Delta)&:=&\{f_{\cal O}:\hat A\rightarrow \hat B
    \mid {\rm Prob}(B\in f(\Delta);\rho)=1\}\nonumber\\[2pt]
        &\,=&\{f_{\cal O}:\hat A\rightarrow \hat B
                \mid {\rm tr}(\rho\,\hat E[B\in f(\Delta)])=1\}.
                                         \label{Def:nurho}
\end{eqnarray}

\section{Conclusion}
In this paper we have suggested that
topos-theoretic notions, in particular the idea of
a topos of presheaves on a base-category $\cal C$
of suitably chosen `contexts', may well be useful
both in the foundations of quantum theory, and in
quantum gravity. Of course, much remains to be
done in applying these notions to the research
areas listed in Section 3. Even for the last area
listed (Section 3.7), where the application has
been worked out in detail, there are many further
natural questions to investigate: for example, one
might consider how standard topics such as the
uncertainty relations or non-locality appear in
this framework. But we hope that such an
open-ended situation---even in the area of
logical, rather than physical, theorizing---might
please someone so active and enthusiastic about
logical and physical research as Marisa Dalla
Chiara!

\bigskip\noindent
{\bf Acknowledgements}.

\noindent JNB thanks the Newton Institute,
Cambridge, England for hospitality during the
preparation of this paper. CJI thanks Ntina
Savvidou, Lee Smolin, and Fotini Markopoulou for
useful discussions.


\begin{thebibliography}{10}

\bibitem{IB98}
C.J. Isham and J.~Butterfield.
\newblock A topos perspective on the {K}ochen-{S}pecker theorem:
{I.} {Q}uantum states as generalised valuations.
\newblock {\em Int.\ J.\ Theor.\ Phys.}, {\bf 37},
2669--2733, 1998.

\bibitem{IB99}
J.~Butterfield and C.J.~Isham.
\newblock A topos perspective on the {K}ochen-{S}pecker theorem:
{II.} {C}onceptual aspects, and classical
analogues.
\newblock {\em Int.\ J.\ Theor\ Phys.}, {\bf 38}, 827--859, 1999.


\bibitem{HIB99}
J.~Hamilton, C.J.~Isham and J.~Butterfield.
\newblock A topos perspective on the {K}ochen-{S}pecker theorem:
{III.} Von Neumann algebras as the base category.
\newblock In preparation, 1999.


\bibitem{KS67}
S.~Kochen, and E.~Specker.
\newblock The problem of hidden variables in quantum mechanics.
\newblock {\em J.\ Math.\ and Mech.} {\bf 17}, 59-87, 1967.



\bibitem{BVN36}
G.~Birkhoff, and J.~von Neumann.
\newblock The logic of quantum mechanics.
\newblock {\em Ann.\ Math.} {\bf 37}, 823-843, 1936.



\bibitem{DCG99}
M.L.~Dalla Chiara, and R.~Giuntini.
\newblock Quantum Logics, in D.~Gabbay and F.~Guenthner
(eds), {\em Handbook of Philosophical Logic}.
\newblock  Kluwer, Dordrecht, to appear.


\bibitem{DCG94}
M.L.~Dalla Chiara, and R.~Giuntini.
\newblock Unsharp quantum logics.
\newblock {\em Found.\ Physics} {\bf 24}, 1161-1177, 1994.


\bibitem{CL94}
G.~Cattaneo, and F.~Laudisa.
\newblock Axiomatic unsharp quantum logics.
\newblock {\em Found.\ Physics} {\bf 24}, 631-684, 1994.


\bibitem{BI99}
J.~Butterfield, and C.J.~Isham.
\newblock Spacetime and the philosophical challenge of quantum gravity, in C.~Callender and N.~Huggett
(eds), {\em Physics meets Philosophy at the Planck
Scale}.
\newblock  Cambridge University Press, forthcoming.
\newblock gr-qc 9903072


\bibitem{Far75}
M.~Farrukh.
\newblock Application of nonstandard analysis to quantum mechanics.
\newblock {\em J.\ Math.\ Physics} {\bf 16}, 177-200, 1975.


\bibitem{Lav96}
R.~Lavendhomme.
\newblock {\em Basic Concepts of Synthetic Differential
Geometry}.
\newblock Kluwer, Dordrecht, 1996.


\bibitem{Sav98}
K.~Savvidou.
\newblock The action operator for continuous-time histories.
\newblock {\em J.\ Math.\ Physics}, forthcoming.
\newblock gr-qc 9811078



\bibitem{Mar98}
\newblock F.~Markopoulou.
\newblock The internal description of a causal set: What the universe looks like from the inside.
\newblock To appear.
\newblock gr-qc 9811053



\bibitem{I94}
\newblock C.J.~Isham.
\newblock Quantum logic and the consistent histories approach
 to quantum theory.
\newblock {\em J.\ Math.\ Phys.}, {\bf 23}, 2157-2185, 1994.

\end{thebibliography}
\end{document}